\documentclass[aps,prb,preprint,showpacs]{revtex4}

\begin{document}
\title{Anomalous refractive properties of a two-dimensional photonic band-gap prism}
\author{J. Bravo-Abad, T. Ochiai\cite{address} and J. S\'{a}nchez-Dehesa\cite{email}}
\affiliation{Departamento de F\'{\i}sica Te\'{o}rica de la Materia Condensada, Facultad de
Ciencias (C-V), \\
Universidad Aut\'onoma de Madrid, Madrid 28049, Spain}

\begin{abstract}
An analysis of the optical response of a triangular-shaped photonic band-gap
prism is presented. Numerical simulations have been performed in the
framework of multiple-scattering theory, which is applied considering spot
illumination to avoid diffraction effects. First of all, refractive
properties in the frequency range below the first TM band-gap are analyzed
and compared with the available experimental data. It validates the approach
employed and supports the predictions obtained in the frequency range above
the gap. At these high frequencies we found an unusual superprism effect
characterized by an angle- and frequency-sensitivity of the intensity
of outgoing beams. We report several representative examples that could be
used in device applications. The results are interpreted in terms of the
corresponding semi-infinite photonic crystal, through the analysis of the
coupling between external radiation and bulk eigenmodes, using the 2D Layer-%
Korringa-Kohn-Rostoker method. The procedure presented here constitutes a
simple but functional alternative to the methods used until now with the same
purpose.
\end{abstract}

\maketitle
\section{Introduction}

Since the seminal works of Yablonovitch \cite{Yablonovitch} and John \cite
{John}, photonic crystals (PCs) have generated a great interest, both in
fundamental and applied physics. These materials could have an important
role in the development of the photonic information technology. This is
especially true in the case of two-dimensional (2D) PCs due to their easy
fabrication by standard lithographic techniques. On the other hand, recent
advances in growing techniques are also allowing the fabrication of planar
devices based on three-dimensional (3D) systems\cite{SMYang}.

The study of optical properties of PC using numerical tools \cite
{Joannopoulos,Sakoda,Soukoulis,Busch} is an essential procedure in order to
further the understanding of the physical properties of PCs. The
potential applications of PCs has motivated the development of efficient
theoretical methods that allow us to obtain realistic comparisons between
theory and experiments. One of the most versatile methods is the Multiple
Scattering Method (MSM). Its formalism is based on the expansion of the 2D
(3D) radiation field in cylindrical (spherical) harmonics centered at each
scattering object, and it has been successfully applied in the analysis of
interaction of radiation with dielectric and metallic finite 2D and 3D
structures \cite
{Felbacq,Tayeb,Li1,Li2,Chow,Yonekura,Nojima,Tetsu,Moreno,FJGarcia}.

Several authors have analyzed the optical response of structures based on
2D-PCs\cite{Notomi,Halevi,Enoch,Pendry,Chung}. Recently, 2D-PCs with
non-parallel boundaries have been investigated both numerically, by using
the finite-difference-time-domain (FDTD) method \cite{Baba}, and
experimentally \cite{Wu}. However, a complete theoretical analysis including
the study of the coupling between incident radiation and bulk eigenmodes is
still lacking.

The goal of this work is to present a comprehensive analysis of the optical
properties of a photonic bandgap prism by using the Multiple Scattering
formalism and other numerical techniques.  The prism, which has a triangular
shape, consists of a hexagonal array of non-overlapping dielectric cylinders
of infinite height. The analysis is made in the low frequency region (below
the first TM band-gap), where a comparison with available experimental data
\cite{Lin} is presented, and also in the frequency domain above the first TM
band-gap, where anomalous refractive properties are predicted. The results
of the MSM simulations are interpreted in terms of the corresponding
semi-infinite PC. Moreover, in order to get a physical insight of the
optical response in this structure, we also have made quantitative
predictions of the coupling strength between the incident light and the bulk
eigenmodes that can be excited; i.e., the eigenmodes that satisfy the
kinetic matching conditions at the air/PC interface. A complete theoretical
analysis by means of a combination of plane wave expansion (PWE) and
2D-Layer-Korringa-Kohn-Rostoker (2D-LKKR) method, has allows us to get
accurate predictions which could be confirmed experimentally. The good
agreement found between theoretical results and available experiments 
supports the different approaches employed.

In our studies we have found an unusual superprism phenomenon, in which
the intensity of outgoing beams shows a dependence on both the frequency and
incident angle of the external beam. This differs from the superprism effect
reported until now\cite{Kosaka}, where large dependence is found only in outgoing
beams directions. Also, it must be emphasized that this work provides
simple but working guidelines to design functionalities of
super-prism structures based on a 2D photonic band-gap (PBG) systems. On the
other hand, this work also shows that the MSM is competitive with the
time-demanding FDTD method in order to simulate actual 2D structures.

The paper is organized as follows. In Section II we describe briefly the
basis of the MSM employed to study the optical response of the prism. The
results of numerical simulations and the discussion in terms of the
semi-infinite model is presented in Section III. Finally, in Section IV we
summarize the work.

\section{Numerical method}

Here, the MSM is applied to study a finite lattice of dielectric cylinders
(of dielectric constant $\epsilon _{r}$) infinitely height embedded in a
background of different dielectric constant ($\epsilon _{b}$). For a
complete description of the numerical algorithm employed see, for example,
the paper by Ochiai and S\'{a}nchez-Dehesa\cite{Tetsu} and references
therein. In brief, the numerical simulations are based on the solution of
the following self-consistent equation \cite
{Felbacq,Tayeb,Li1,Li2,Chow,Yonekura,Nojima,Tetsu,Moreno,FJGarcia}:

\begin{equation}
\widetilde{\psi }_{\alpha }^{{\rm ind}}=t_{\alpha }(\widetilde{\psi }%
_{\alpha }^{{\rm ext}}+\sum_{\beta \neq \alpha }G_{\alpha \beta }\widetilde{%
\psi }_{\beta }^{{\rm ind}}),  \label{basic}
\end{equation}
where the field $\psi =E_{z}(H_{z})$ for TM (TE) polarization. $\widetilde{%
\psi }_{\alpha }^{{\rm ind}}$ in Eq.(\ref{basic}) is a column vector that
contains the components of the multipole expansion of the radiation field
induced by the cylinder $\alpha $ centered at ${\bf r}_{\alpha }$, $%
\widetilde{\psi }_{\alpha }^{{\rm ext}}$ represents the external wave, $%
t_{\alpha }$ is the t-matrix, and $G_{\alpha \beta }$ is the propagator from
cylinder $\beta $ to $\alpha $.

Most of the previous simulations using the 2D-MSM employed an incident wave
having a plane wavefront as external source . However, in the case under
study here, a structure with non-parallel boundaries, it is necessary to use
spot illumination in order to avoid diffraction effects. Moreover, spot
illumination is essential to make comparisons with available experiments\cite
{Lin}, which use collimated beams in the microwave regime. So far, to the
best of our knowledge, there have been two different procedures to simulate a
collimated beam: (1) the first one defines a finite aperture in front of the
sample \cite{Li1,Li2}. (2) the second one simulates the spot illumination by
using a Gaussian beam \cite{Enoch,Gralak,Pelosi}. We have demonstrated that,
for our system, a Gaussian beam is more appropriate because it reproduces
the measurements fairly well.

The incident Gaussian beam of frequency $\omega$, focusing at ${\bf r}_0$,
can be expressed as the following very directional solution of the
homogeneous Helmholtz equation\cite{Nieto-Vesperinas}:

\begin{equation}
\psi ^{{\rm ext}}({\bf r})=\int_{-k_{0}}^{k_{0}}dk_{y}e^{-Dk_{y}^{2}}e^{i%
{\bf k}({\bf r}-{\bf r}_{0})},  \label{incident1}
\end{equation}
where ${\bf r}$ is the position vector, which in polar coordinates ${\bf r}%
=(r,\theta )$. ${\bf k}$ is the wave vector of a plane wave in the beam, $%
{\bf k}=(k_{x},k_{y})$, where $k_{x}=\sqrt{k_{0}^{2}-k_{y}^{2}}$, and $%
k_{0}=\omega \sqrt{\epsilon _{b}}/c$. Parameter $D$ determines the width of
the beam.

Therefore, the $\ell $-th component of the external field is given as:

\begin{equation}
\psi _{\alpha ,\ell }^{{\rm ext}}=i^{\ell }k_{0}\int_{-\frac{\pi }{2}}^{%
\frac{\pi }{2}}d\theta \cos {\theta }e^{g_{1}(\theta )+ig_{2}(\theta )},
\label{incident2}
\end{equation}
where

\begin{eqnarray}
\lefteqn{g_{1}(\theta )=-D\:k_{0}^{2} \: \sin ^{2}\theta } \\
g_{2}(\theta ) &=&k_{0}\:\{r_{\alpha }\cos {(\theta +\theta _{inc}-\theta
_{\alpha })}-r_{0}\cos {(\theta +\theta _{inc}-\theta _{0})}\}-\ell \: \theta
\end{eqnarray}
and $(r_{\alpha },\theta _{\alpha })$ and $(r_{0},\theta _{0})$ are the
polar coordinates of the $\alpha $-th cylinder and the focusing point,
respectively.

We are interested in the far field response of the structure. It can be
easily found that in the far field regime \cite{Sheng}:

\begin{equation}
\psi^{ext}({\bf r}) \rightarrow \left\{ 
\begin{array}{ll}
a_{0}( \theta )\:e^{ik_{0}r} /\sqrt{r} & 
\mbox{if $-\frac{\pi}{2} \leq \theta
\leq \frac{\pi}{2}$} \\ 
a_{0}(\theta-\pi)\:e^{-ik_{0}r} /\sqrt{r} & \mbox{otherwise,}
\end{array}
\right.
\end{equation}
where $a_{0}(\theta )=\sqrt{2\pi k_{0}}\:e^{-i\frac{\pi }{4}}\:e^{-\text{D}%
k_{0}^{2}\sin ^{2}\theta }\cos {\theta }$.

The magnitude of interest in this problem is the intensity at infinity ($%
I_{s}$). It is defined by the Poynting vector flux per unit angle calculated
at far field:

\begin{equation}
I_{s}(\theta )=\left\{ 
\begin{array}{ll}
|a_{0}(\theta )+f_{s}(\theta )|^{2} & 
\mbox{if $-\frac{\pi}{2} \leq \theta
\leq \frac{\pi}{2}$} \\ 
|f_{s}(\theta )|^{2}-|a_{0}(\theta )|^{2} & \mbox{otherwise,}
\end{array}
\right. 
\end{equation}
where $f_{s}(\theta )$ is the scattering amplitude, which is obtained from
Eq.(\ref{basic}) as a linear combination of $\psi _{\alpha ,\ell }^{{\rm ind}%
}$ (Ref. 15).

\section{Results and discussion}

The optical response to TM-polarized light of a 2D photonic band-gap prism
based on a hexagonal array of 1540 alumina rods ($\epsilon _{r}=$8.9) in
air ($\epsilon _{b}=$1.0) is studied. In order to compare with the
experiments performed by Lin {\it et al.}\cite{Lin} the radius of cylinders
is set to $r=$0.188 $a$, where $a$ is the lattice parameter. The external
shape of the prism is triangular: the three boundaries of the prism are
composed of 55 cylinders each one, and they are parallel to the $\Gamma K$
direction. A scheme of the structure and the definitions of incident angle ($%
\theta _{inc}$) and deviation angle ($\delta $) are shown in Fig. 1. The
system of reference used along the calculations is also plotted.

The photonic band structure for TM-polarized modes of the corresponding
infinite lattice, which has been calculated by means of a plane wave
expansion\cite{Johnson}, is plotted in Fig. 2, where the frequencies are
given in reduced units. A factor of 2$\pi c/a$ (c is the light velocity in
air) must be used to obtain the absolute values. So, in the rest of the
paper the frequency is given in reduced units. Results in Fig. 2 indicate
that a complete band-gap exists between $\omega _{L}=$0.325 and $\omega
_{U}= $0.497.

First of all, frequencies inside the band-gap were examined in order to
check whether the bulk properties of the hexagonal lattice were already achieved.
Let us consider a frequency slightly greater than $\omega _{L}$, for
instance, 0.33. The electric pattern computed for the case of normal
incidence ($\theta _{inc}=$0$^{\text{o}}$), $D$ $=$20 $a^{2}$, and ${\bf {r_0}%
}$=0 is visualized in Fig. 3. It can be seen that no light propagates inside
the prism and, therefore, this frequency is inside the photonic band-gap,
which is in agreement with the photonic band structure shown in Fig. 2.

In what follows we report the optical response in two different regimes,
below and above the first photonic TM band-gap. We obtained converged
results by considering $\ell =1$ as the maximum angular momentum for
frequencies below the band-gap, and $\ell =2$ for the frequencies considered
above the band-gap.

\subsection{Behavior at frequencies below the first TM band-gap}

In the very low frequency region the equifrequencial curves are circles and,
therefore, it is possible to define an effective refraction index
independent of incident angle \cite{Notomi}.

It can be stated that, by using the MSM and for frequencies below the first
TM band-gap, it is possible to accurately find the deviation angle of the
beam transmitted across the prism only by assuming certain ${\bf {r}_{0}\ne}$%
0 and $D$ values large enough\cite{footnote1}.

In addition, the asymmetry observed in the transmitted peaks can be
explained by the Fresnel factor because of the fact that the incident beam
is represented by superposition of plane waves with different incident
angles and different weights [Eq.(2)].

A procedure to get the effective refractive index ($n_{eff}$) from the MSM
is to fit the numerical results $(\theta _{inc},\delta )$ to the expression
obtained applying Snell's law in the case of non-parallel boundaries
structure with a homogeneous $n_{eff}$:

\begin{equation}
\delta =\theta _{inc}+\sin ^{-1}((sin\alpha )(n_{eff}^{2}-\sin ^{2}\theta
_{inc})^{\frac{1}{2}}\mbox{}-\sin {\theta _{inc}}\cos \alpha )-\alpha ,
\label{neff}
\end{equation}
where $\alpha$ is the prism angle, which is 60$^{\text{o}}$ for the
triangular prism under analysis. Figure 4 shows the dependence of $\delta$
on $\theta _{inc}$ for several angles at $\omega$=0.26 (black dots). The
solid line corresponds to a fit of MSM data to Eq.(\ref{neff}). From this
fit $n_{eff}=$1.54 is obtained. Experimentally, Lin {\it et al.}\cite{Lin}
reported $n_{eff}\approx$1.57 (dashed line in Fig. 4). Therefore, our
simulations based on the MSM confirms that at this low frequency ($\omega$%
=0.26) the refractive index does not depends on $\theta_{inc}$ and,
moreover, its value fairly agrees with experimental data (about 2$\%$ of
relative error).

The previous considerations allow us to analyze theoretically the optical
response for all frequencies below the first band gap. Unfortunately, we
found that the contribution of the transmitted beam for frequencies near the
band edge are not clearly observed in the calculations (its transmission is
almost zero). In other words, the scattered intensity is mainly determined
by the shape of the prism. The result is summarized in Fig. 5(a), where $%
n_{eff}$ obtained from the MSM simulations (black dots) are compared with
experimental data (white circles) and with a simplified model (solid line)
that considers light propagation across single air/PC interfaces (see
Appendix A). The last approach predicts a sharp decreasing of $n_{eff}$ near
the band edge, which, unfortunately, is confirmed neither by
the MSM nor by the experimental set up due to their intrinsic limitations.
Figure 5(b) shows the behavior of $n_{eff}$ as a function of $\theta _{inc}$%
. It can be observed that, as it is expected, $n_{eff}$ does not depend on $%
\theta _{inc}$ at low frequencies, but near the band edge, the critical
frequency at which $n_{eff}$ starts to decrease depends on $\theta _{inc}$.
This phenomenon is a result of the shape of the equifrequencial curves near
the band edges. This prediction should stimulate more accurate experimental
works looking for its demonstration.

Finally, let us remark that small discrepancies observed between
experimental and numerical results can be mainly attributed to the fact that
the actual beam used experimentally is not well enough simulated by the 2D
numerical beam introduced in the simulations by the MSM. In other words, the
spot size and collimation of a 2D Gaussian beam cannot be controlled to
reproduce accurately the properties of the actual beam. This is the main
drawback of the simulations based on the MSM, which produces fictitious
diffraction effects. Fortunately, for higher frequencies these effects
disappear, and it is possible to characterize interesting phenomena as it is
reported in what follows.

\subsection{Behavior above the first TM band-gap}

In order to study the optical response in frequencies above the first
complete TM band-gap, we have started by analyzing the case of infinite
structures and have chosen the ranges of frequency and incident angle at
which anomalous phenomena are expected when light crosses an air/PC
interface. Equifrequencial curves between 0.509 and 0.627 are plotted in
Fig. 6(a). Notice that near the curves associated to frequency 0.537 [thick
lines in Fig. 6(a)] the equifrequencial curves consist of two separated
regions: a hexagonal shaped region (centered at $\Gamma $ point) and a
triangular shaped region (centered at K point). Therefore, it is expected
that such anisotropic equifrequencial regions lead to anomalous response of
a finite structure based on the same lattice periodicity.

The procedure described in Appendix A allows us to compute the propagating
angle ($\theta _{pro}$) inside the PC as a function of the incident angle for a
beam crossing the air/PC interface at $\omega =$0.537. The result of this
calculation, which is depicted in Fig. 6(b), shows that above a critical
angle  ($\theta_{c}$) the incident beam splits into two. They correspond to
the curves of frequency 0.537 which have group velocity with positive $y-$%
component. The great dependence of $\theta _{pro}$ on $\theta _{inc}$ of
these solutions from 20$^{\text{o}}$ to 25$^{\text{o}}$ degrees and from 25$%
^{\text{o}}$ to 40$^{\text{o}}$ is associated to the rounded corners of the
hexagonal- and triangular-shaped regions, respectively.

A comprehensive analysis of the scattering properties in the frequency
range between 0.5 and 0.6 and for incident angles between 20$^{\text{o}}$
and 40$^{\text{o}}$ has been performed by using the MSM. This analysis has
allowed us to select the most interesting phenomena.

Figure 7(a) shows the intensity map as a function of $\delta $ and $\theta
_{inc}$ computed with $\omega $=0.537. The parameters of the Gaussian beam
employed in both simulations\cite{footnote2} are $D=$30$a^{2}$ and $y_{0}$=10%
$a$ . As can be seen from this figure, we do not obtain a great dependence
of $\delta $ as a function of $\theta _{inc}$ as could be expected in the
usual definition of superprism effect. Instead, we obtain an angular
sensitivity of the transmitted beam intensities. In order to prove clearly
this anomalous effect, Fig. 7(b) shows the comparison between intensity
spectrum corresponding to the incident angle with largest transmission ($%
\theta _{inc}$=22.8$^{\text{o}}$), and the spectra corresponding to incident
angles that differ in 3.2$^{\text{o}}$. As it is shown in this figure, a
small change in $\theta _{inc}$ implies a dramatic fall in the transmittance.

In addition, it is interesting to know the dependence of $I_{s}$ on $\omega $
and $\delta $ for $\theta _{inc}$=22.8$^{\text{o}}$. The result of this
calculation is shown in Fig. 8(a). This second transmitted beam
only appears in a very narrow range of frequencies; i.e., from 0.533 up to
0.582 approximately. In order to show quantitatively this effect, Figure 8(b)
shows a sharp change in the intensity features of the spectra for very small
variations (about 0.7 $\%$) of frequency. Again we obtain a novel refractive
effect characterized by the large frequency-sensitivity in the intensity of
the outgoing beams.

The phenomena mentioned previously suggest the above configuration as a
sensitive beam splitter, both in angle- and frequency regime.

A similar analysis to the one described previously was carried out for
another frequency of interest, $\omega $=0.539. The corresponding intensity
map, with $\theta _{inc}$ and $\delta $ as variables, is plotted in Fig.
9(a). From this figure we have selected the angle $\theta _{inc}$=40$^{\text{%
o}}$ as the most interesting case. Figure 9(b) shows its corresponding
intensity spectrum compared with the spectra associated to two angles
differing in 5.2$^{\text{o}}$. As can be seen from this figure, the main
transmitted beam found at $\theta _{inc}$=40$^{\text{o}}$ disappears
completely for $\theta _{inc}$=34.8$^{\text{o}}$, while it is almost the
same for $\theta _{inc}$=45.2$^{\text{o}}$. 

Also we have made a frequency analysis by fixing $\theta _{inc}$ at 40$^{%
\text{o}}$. The results are summarized in Fig. 10(a), where it is seen that
the main transmitted beam only exists in two separated frequency regions.
Particularly, Fig. 10(b) shows the behavior in the region centered at $%
\omega $=0.539. Notice that a change in frequency of about 4$\%$ implies a
sharp decrease in the transmission of the incident radiation through the
prism. Therefore, this configuration could be suitable for the design of a
PC filter for a short range of frequencies. Similar conclusion could be
obtained for frequencies above 0.58, where the prism is almost transparent
and the transmitted beam undergoes a small deviation angle ($\delta \approx
-20^{o}$).

Physical insight of the mechanism originating the phenomena observed in
Figs. 7, 8, 9, and 10, can be obtained from the 2D-LKKR method
\cite{Ohtaka,Moroz}. In fact, we have implemented a 2D-version of the
procedure proposed by two of us in a previous work \cite{Tetsu2} (see
Appendix B). In this way, it is possible to obtain the branching ratio (BR) of
the different bulk eigenstates of the PC that can couple to the incident
light.

In the following, we will refer to the configurations defined by ($\omega $, 
$\theta _{inc}$)=(0.537, 22.8$^{\text{o}}$) and ($\omega $, $\theta _{inc}$%
)=(0.539, 40.0$^{\text{o}}$) as case 1 and case 2, respectively.

In case 1 (see Fig. 11), simulations using the MSM predict that in
addition to the reflected beam (R) at the front air/PC interface (133.27$^{%
\text{o}}$), two transmitted beams appear: one of them at -105.82$^{\text{o}}
$ (T$_{1}$) and another at 14.42$^{\text{o}}$(T$_{2}$). In this case, if we
apply the kinetic matching conditions to the frontal air/PC interface, we obtain
that only one eigenstate with group velocity inward the PC couples to the
incident light. The perpendicular component, $k_{\perp }$, of this
eigenstate wave vector (${\bf k}_{pro}$) to the boundary is -0.3592 (in
units of $2\pi /a$) [see inset in Fig. 11(b)]. A similar analysis can be
performed at boundary 2 in Fig. 11(a). Since boundaries 1 and 2 are not
parallel, it is necessary to make a 60$^{\text{o}}$ rotation of ${\bf k}%
_{pro}$ before applying the matching conditions. As a result of this
analysis, we conclude that an output beam should appear at $\delta =$-105.48$%
^{\text{o}}$, in good agreement with the MSM-based simulations. Also we have
to consider the reflection in boundary 2 and the following refraction in
boundary 3. For this path the predicted deviation angle is 14.4$^{\text{o}}$%
, and we expect small intensity for this peak since it comes from considering
two refractions and one reflection. On the other hand, from the 2D-LKKR it
is possible to obtain the reflectance of the corresponding semi-infinite PC
at this incident angle. The reflectance is about 0.544. This result is
supported by the MSM, where the intensity of the peak associated with the 
reflected beam (R in Fig. 11) is related to the intensity of the incident beam
by that factor.

The field pattern and transmission spectrum obtained from the MSM-based
simulations for case 2 ($\omega =$0.539 and $\theta _{inc}=$40$^{\text{o}}$)
are plotted in Figs. 12(a) and 12(b), respectively. A very small
reflected peak (R) appears at 101.16$^{\text{o}}$. The main transmitted beam
(T$_{1}$) with negative refraction index appears at -146.16$^{\text{o}}$,
while two secondary transmitted beams are seen at -93.24 (T$_{2}$) and -19.98%
$^{\text{o}}$ (T$_{3}$). This result can be understood in similar terms to
the preceding case. In other words, by analyzing the light transmission
through single interfaces. Now we have two eigenstates satisfying the
kinetic matching conditions at the frontal interface; their $k_{\perp }$
values inside the PC, in units of $2\pi /a$, are 0.5095 and -0.1237 [see
inset in Fig. 12(b)], and the corresponding BRs are 0.923 and 0.077,
respectively. The reflectance of the corresponding semi-infinite PC is about
0.085. This value of the reflectance is in accordance with the small
intensity peak of the reflected beam in comparison with the main transmitted
beam. The matching conditions applied to boundary 2 in Fig. 12(b) concludes
that the first outgoing beam should appear at $\delta =$-145.75$^{\text{o}}$,
in good agreement with the MSM results. If we also take into account the
beam with negative $k_{\perp }$, we obtain that a second outgoing beam
appears at $\delta =$-93.0$^{\text{o}}$ (also in agreement with Fig. 12). The
great difference between the two intensities of these two refracted beams
could be estimated using the BR values. Moreover, for the path corresponding
to the light reflected at the boundary 2 and refracted at boundary 3, the
predicted deviation angle is -20.0$^{\text{o}}$ and also we expect a
secondary refracted beam, since the dominant eigenstate in boundary 2 is the
refraction.

Table I summarizes the results regarding the directionality of the scattered
light. It demonstrates that the deviation angle of the transmitted light
through non-parallel boundaries can be predicted by a simplified model based
on the light transmission across semi-infinite air/PC interfaces. Its
results agree fairly well with the complete simulation of the finite system
performed by MSM-based simulations. In other words, the combination of the
Hellmann-Feynmann theorem in the plane wave expansion together with the
2D-LKKR method is enough to analyze the optical response of this kind of
structures in the high frequency region.

\begin{table}[htb]
\caption{Comparison of the deviation angles (in degrees) of beams transmitted
across the prism. The results under the columns 'Model' are obtained by the
procedure described in Appendix A.}
\begin{ruledtabular}
\begin{tabular}{lcccc}
& \multicolumn{2}{c}{$\omega$=0.537, $\theta_{inc}$=22.8$^{\text{o}}$} & 
\multicolumn{2}{c}{$\omega$=0.539, $\theta_{inc}$=40$^{\text{o}}$} \\ 
& MSM & Model & MSM & Model \\ 
\colrule $T_{1}$ beam & -105.82 & -105.48 & -146.16 & -145.75 \\ 
$T_{2}$ beam & 14.42 & 14.40 & -93.28 & -93.00 \\ 
$T_{3}$ beam & --- & --- & -19.98 & -20.00
\end{tabular}
\end{ruledtabular}
\end{table}

\section{summary}

In this work we have analyzed the optical response of a triangular prism
made of a hexagonal array of dielectric cylinders in air. The approach
employed, based on the Multiple Scattering formalism, has been supported by
the available experimental data existing at frequencies below the first TM
band-gap. Also, we have predicted anomalous superprism phenomena above the
first TM band-gap. A comprehensive analysis in this high frequency region
has been performed by combining the PWE method and the 2D-LKKR method. We
have demonstrated that, if the prism is large enough (diffraction
effects are negligible), a procedure based in the analysis of the
corresponding semi-infinite system agrees very well with the simulation of
the complete finite system by the MSM. We have described two cases with
possible device application. The results presented in this work support the
applied procedure as an efficient numerical tool in the understanding of the
optical response in 2D-PC structures with non-parallel boundaries. We hope
that this work motivates further experiments that confirm its predictions.
Finally, the procedure here employed has been also extended to study
analogously 3D prisms composed of PCs having a strong anisotropy in its
photonic dispersion relation, and the results will be presented in  a
forthcoming publication.

\begin{acknowledgments}
The authors would like to thank F.J. Garc\'{\i}a de Abajo for helpful
discussions. This work was supported by European Commission, Project No.
IST-1999-19009 PHOBOS and by the Spanish CICyT Project No.
MAT2000-1670-C04-04. We also acknowledge the computing facilities provided
by the Centro de Computaci\'{o}n Cient\'{\i}fica at Universidad Aut\'{o}noma
de Madrid. One of the authors (J.B-A.) thanks F. L\'{o}pez-Tejeira for
valuable discussions.
\end{acknowledgments}

\appendix

\section{light transmission across a single air/PC interface}

Let us assume a semi-infinite 2D PC defined in the region $y>$0, and air in
the another semi-plane. In order to know which eigenstates of the PC can
be excited by an external light, it is important to project the bulk photonic
band structure on the surface Brillouin zone (SBZ). A typical projected band
diagram is shown in Fig. 13. This figure represents the band structure
projected on the $\Gamma K$ boundary surface, which is the common boundary
surface of the triangular prism under study. In Fig. 13 the shaded regions
define the allowed bulk eigenstates and the thick black line defines the
light line. This line defines the critical frequency above which external
plane waves are allowed to exist. The white regions represent pseudogaps in
the dispersion relation. Therefore, external beams having frequencies in a
pseudogap cannot enter into the PC, as it is shown in Fig. 3.

Another criterion of coupling arises from the group velocity of the bulk
eigenmodes. Only the bulk eigenstates with positive $v_{y}$ can couple with
the incident light. Those eigenstates with $v_{y}<0$ should be physically
neglected.

In summary, the kinetic matching condition between the bulk eigenmodes and
the incident light are the following: (1) The frequency and the reduced wave
vector in the SBZ are conserved, (2) The bulk eigenmodes are above the light
line, and (3) $v_{y\text{ }}$is positive. Once these conditions are
satisfied, the incident angles $\theta _{inc}$ and the propagating angles $%
\theta _{pro}$, which are given by 
\begin{eqnarray}
k &=&\frac{\omega }{c}(\sin \theta _{inc},\cos \theta _{inc}), \\
v &=&|v|(\sin \theta _{pro},\cos \theta _{pro}),
\end{eqnarray}
can be related through an effective refraction index ($n_{eff}$) defined as:

\begin{equation}
sin(\theta _{inc})=n_{eff}{\sin \left( \arctan \left( \frac{v_{x}}{v_{y}\ }%
\right) \right) },  \label{neff2}
\end{equation}
where $(v_{x},v_{y})$ are the components of the group velocity, which are
computed directly from the dispersion relations by using the
Hellmann-Feynman theorem.

\section{branching ratio}

Figure 2 shows that at high frequencies several eigenmodes can be
kinetically matched to the external light. Hence, it is necessary to know
quantitatively what the coupling strength is between the incident light and
the bulk eigenstates of the 2D PC. Here, the 2D layer-KKR method is used for
that purpose in a similar manner as was employed its 3D counterpart in the
analysis of the superprism effect in opals\cite{Tetsu2}.

We have considered a slab large enough for the deepest region in the
slab to be regarded as a superposition of the actual bulk eigenstates $%
\alpha $; i.e., the field at this center region is assumed not be affected
by the finite size of the slab. In the 2D-LKKR method three different
regions can be clearly separated: (1) Left region, defined by the scattering
matrix $Q_{L}$. (2) Center or deepest region. (3) And right region, whose
scattering matrix is $Q_{R}$. If we consider that a plane wave impinges the
slab from the left, the following relationships are satisfied \cite
{Tetsu2,Ohtaka}:

\begin{equation}
\left( 
\begin{array}{c}
{\bf E}_{d}^+ \\ 
{\bf E}_{ref}
\end{array}
\right) = \left( 
\begin{array}{cc}
Q_{++}^L & Q_{+-}^L \\ 
Q_{--}^L & Q_{-+}^L
\end{array}
\right) \left( 
\begin{array}{c}
{\bf E}_{inc} \\ 
{\bf E}_d^{-}
\end{array}
\right) ,  \label{a1}
\end{equation}

\begin{equation}
\left( 
\begin{array}{c}
{\bf E}_{tr}^{+} \\ 
{\bf E}_{d}^{-}
\end{array}
\right) =\left( 
\begin{array}{cc}
Q_{++}^{R} & Q_{+-}^{R} \\ 
Q_{--}^{R} & Q_{-+}^{R}
\end{array}
\right) \left( 
\begin{array}{c}
{\bf E}_{d}^{+} \\ 
0
\end{array}
\right) ,  \label{a2}
\end{equation}
where ${\bf E}_{inc}$, ${\bf E}_{ref}$, ${\bf E}_{tr}$ are column vectors
which contain the Fourier components of the incident, reflected, transmitted
fields, respectively. ${\bf E}_{d}$ corresponds to the field in the inner
void region of the slab, '$+$' denotes propagation from left to right and '$%
- $' from right to left.

From Eq.(\ref{a1}) and (\ref{a2}) it is easy to obtain:

\begin{eqnarray}
{\bf E}_{d}^+=(1-Q_{+-}^{L}Q_{-+}^{R})^{-1} \: Q_{++}^{L} \: {\bf E}_{inc} \\
{\bf E}_{d}^-=Q_{-+}^R \: (1-Q_{+-}^{L}Q_{-+}^{R})^{-1} \: Q_{++}^{L} \: 
{\bf E}_{inc}
\end{eqnarray}

Let us consider the $\alpha $-th left eigenstate (${\bf V}_{\alpha ,L}^{\pm }$)
of the transfer matrix which relates the Fourier components of the electric
between two consecutive voids in a infinite PC. It is possible to define the
projection of the eigenstate in the deepest void of the slab in the true $%
\alpha $-th eigenstate through the coefficients $c_{\alpha }$ defined by:

\begin{equation}
c_{\alpha}=({\bf V}_{\alpha,L}^+)^* \: {\bf E}_{d}^+ + ({\bf V}%
_{\alpha,L}^-)^* \: {\bf E}_{d}^-
\end{equation}

These coefficients allows to define the BR corresponding to the bulk
eigenstates:

\begin{equation}
BR_{\alpha }=\frac{|c_{\alpha }|^{2}}{\sum_{\alpha ^{\prime }}|c_{\alpha
^{\prime }}|^{2}},
\end{equation}
where the summation is over all the bulk states satisfying the kinetic
matching condition. So, BR$_{\alpha }$ gives an estimation of the coupling
of the external light with the $\alpha $-th bulk eigenstates in the 2D PC,
and it is normalized to the total coupling of eigenstates available for
light transmission.

\newpage

\begin{figure}[H]
\caption{Schematic view of the triangular prism under study and parameter
definitions. The reference system used in the calculation is also plotted.}
\end{figure}

\begin{figure}[H]
\caption{Photonic band structure for TM-polarized modes corresponding to a
hexagonal lattice of dielectric cylinders ($\protect\epsilon_{r}$=8.9) in
air ($\protect\epsilon _b$=1.0). The cylinder radius is $r$=0.188$a$, where $%
a$ is the lattice constant. The shaded region defines the complete band gap.}
\end{figure}

\begin{figure}[H]
\caption{Electric field pattern, $E_z(x,y)$ generated by a Gaussian beam,
with the parameters $D$=20$a^2$, ${\bf {r}_{0}}$=0 and $\protect\omega$%
=0.33, impinging on the prism.}
\end{figure}

\begin{figure}[H]
\caption{Dependence of deviation angle ($\protect\delta$) on incident angle (%
$\protect\theta_{inc}$) for $\protect\omega=$0.26 (solid points). Solid line
is obtained by fitting Eq.(\ref{neff}). $n_{eff}$=1.54 is obtained from this
fit. Dashed line corresponds to experimental results.}
\end{figure}

\begin{figure}[H]
\caption{(a) Behavior of the effective refractive index ($n_{eff}$) as a
function of frequency. Hollow points corresponds to experimental data, solid
points to MSM simulations and solid line to the simplified model (see text).
(b) Dependence of $n_{eff}$ as a function of the frequency for different
incident angles ($\protect\theta_{inc}$) computed by the simplified model.}
\end{figure}

\begin{figure}[H]
\caption{(a) Equifrequencial curves for the photonic TM-bands in Fig.2. Curves
for frequencies between 0.509 and 0.627 are represented. Solid line
corresponds to $\protect\omega$=0.537. (b) Dependence of the propagation
angle on the incident angle considering $\protect\omega=$0.537. Dashed
line is the result of Snell's law calculated by considering an uniform
medium with a spatially averaged refractive index $n_{ave}=1.419$.}
\end{figure}

\begin{figure}[H]
\caption{(a) Intensity map as a function of the deviation angle ($\protect%
\delta$) and the incident angle ($\protect\theta_{inc}$) computed with $%
\protect\omega$=0.537. White corresponds to zero values and black to the
maximum intensity. Dashed line defines the incident angle in which a maximum of
the transmittance is found. (b) Comparison between the intensity spectra
corresponding to the same frequency ($\protect\omega$=0.537) and three
different angles. 'R' labels the peak corresponding to the reflected intensity.
Notice the angular-sensitivity of intensity features.}
\end{figure}

\begin{figure}[H]
\caption{(a) Intensity map as a function of the deviation angle ($\protect%
\delta$) and the frequency ($\protect\omega$) computed with $\protect\theta%
_{inc}$=22.8$^{\text{o}}$. White corresponds to zero values and black to the
maximum intensity. Dashed line shows the frequency
in which a maximum of the transmittance is found. (b) Comparison between the
intensity spectra corresponding to the same incident angle ($\protect\theta%
_{inc}$=22.8$^{\text{o}}$) and three different frequencies. 'R' labels the peak corresponding
to the reflected intensity. Notice the frequency-sensitivity of intensity 
features.}
\end{figure}

\begin{figure}[H]
\caption{(a) Intensity map as a function of the deviation angle ($\protect%
\delta$) and the incident angle ($\protect\theta_{inc}$) computed with $%
\protect\omega$=0.539. White corresponds to zero values and black to the
maximum intensity. (b) Comparison between the intensity spectra
corresponding to the same frequency ($\protect\omega$=0.539) and three
different angles. 'R' labels the peak corresponding to the reflected intensity.
Notice the angular-sensitivity of intensity features between the top and the
center spectra.}
\end{figure}

\begin{figure}[H]
\caption{(a) Intensity map as a function of the deviation angle ($\protect%
\delta$) and the frequency ($\protect\omega$) computed with $\protect\theta%
_{inc}$=40.0$^{\text{o}}$. White corresponds to zero values and black to the
maximum intensity. Dashed line defines the frequency in which a maximum of
the transmittance is found. (b) Comparison between the intensity spectra
corresponding to the same incident angle ($\protect\theta_{inc}$=40.0$^{\text{o}}$) and
three different frequencies. Notice the great frequency-sensitivity of 
intensity features.}
\end{figure}

\begin{figure}[H]
\caption{(a) Electric field pattern and (b) transmission spectra generated
from a Gaussian beam with $\protect\omega=$0.537 and $\protect\theta_{inc}$%
=22.8$^{\text{o}}$. White arrows indicate the direction of the group
velocity computed by the semi-infinite model (see Appendix A). The solutions
to $k_{\perp}$ and the BR value are shown in the inset.}
\end{figure}

\begin{figure}[H]
\caption{(a) Electric field pattern and (b) transmission spectra generated
from a Gaussian beam with $\protect\omega=$0.539 and $\protect\theta_{inc}$%
=40$^{\text{o}}$. White arrows indicate the direction of the group velocity
computed by the semi-infinite model (see Appendix A). The solutions to $%
k_{\perp}$ and the corresponding BRs values are shown in the inset.}
\end{figure}

\begin{figure}[H]
\caption{Projection of the photonic band structure shown in Fig. 2 on the
surface Brillouin zone of the $\Gamma K$ surface. The shaded regions defines
the bulk eigenmodes. The frequencies of the external radiation field are
defined above the light line (black line), $\protect\omega \geq ck_{||}$.}
\end{figure}


\begin{references}
\bibitem{footnote2}  In this frequency range, the result is almost
independent of these parameters since the beam is very collimated and
diffraction effects are negligible.
\end{references}
\end{document}